\documentclass[conference]{IEEEtran}
\IEEEoverridecommandlockouts
\usepackage{cite}
\usepackage{amsmath,amssymb,amsfonts}
\usepackage{algorithmic}
\usepackage{graphicx}
\usepackage{textcomp}
\usepackage{booktabs}
\usepackage{soul, xcolor}
\usepackage{listings}
\usepackage{enumitem}
\usepackage{multirow}
\usepackage{tabularx}
\usepackage{booktabs}
\usepackage[caption=false]{subfig}
\usepackage[normalem]{ulem}

\def\BibTeX{{\rm B\kern-.05em{\sc i\kern-.025em b}\kern-.08em
    T\kern-.1667em\lower.7ex\hbox{E}\kern-.125emX}}
\begin{document}

\title{Engineering and Experimentally Benchmarking Open Source MQTT Broker Implementations}

\author{\IEEEauthorblockN{Jasenka Dizdarevi{\'{c}}, Marc Michalke and Admela Jukan}
\IEEEauthorblockA{\textit{Technische Universit\"at  Braunschweig} \\
Braunschweig, Germany \\
\{j.dizdarevic, m.michalke, a.jukan\}@tu-bs.de}
}

\maketitle

\begin{abstract}
    
    The Message Queuing Telemetry Transport (MQTT) protocol is one of the most widely used IoT protocol solutions. In this work, we are especially interested in open-source MQTT Broker
     implementations (such as Mosquitto, EMQX, RabbitMQ, VerneMQ, and HiveMQ) and their performance in edge networks with various  the parameters of network delay, variance and packet loss. To this end, we engineer a
    network testbed to experimentally benchmark the performance of these implementations in an edge
     computing context with constrained devices. In more detail, we engineer an automated deployment and
     orchestration of the containerized MQTT broker implementations, with support for deployment across
    either moderately powerful AMD64 devices, or more resource constrained ARM64 devices. The proposed
    MQTT implementations are evaluated in terms of performance offset response time and different payload sizes.
    Results showed that the hardware platform used as well as the message size, and the network parameters (latency, packet loss and jitter) have a significant impact on the resulting performance of various broker implementations. All results, software tools and code are fully reproducible and free and open source. 
    \end{abstract}


\section{Introduction}

With the increasing expansion of Internet of Things (IoT) applications, a significant number of
efforts has been centered around the establishment of effective communication patterns, choosing
from various available standards, frameworks, and communication protocols. Over the past years, a
number of application layer protocols, including Message Queuing Telemetry Transport (MQTT),
Constrained Application Protocol (CoAP), Hypertext Transfer Protocol (HTTP), Extensible Messaging
and Presence Protocol (XMPP),  Advanced Message Queuing Protocol (AMQP), and Data Distribution
Service (DDS), have found their use in IoT related system designs and solutions, without a single
standard that could fit all scenarios.

The choice of an application layer communication protocol is particularly challenging when considering IoT edge/cloud computing systems, where previously deployed cloud services (e.g., communication protocol's broker/server instances) can be located at the edge to support latency-sensitive applications \cite{jha2020iotsim, swamy2020empirical}. Here, in addition to considering performance aspects among different protocol solutions, the wide set of available implementations  for each protocol significantly impacts the understanding of their appropriate utilization. Despite its known shortcomings, such as limited scalability of centralized broker architectures or TCP based transport, MQTT remains the currently most widely adopted publish/subscribe protocol.  To us, of special interest are open-source MQTT broker implementations and their integration in edge network eco-systems. While multiple previous studies have addressed various aspects of MQTT performance from the theoretical and implementation perspective \cite{jukan2023lab}, as well as existing comparisons of different MQTT brokers, works that integrate various MQTT broker implementations experimentally in different network scenarios are few and far between. To the best of our knowledge, ours is the first paper to address this practically highly relevant method to performance analysis, especially in the edge networking context.

We focus on performance benchmarks of five well known open source MQTT broker implementations, including Mosquitto, EMQX, RabbitMQ, VerneMQ, and
HiveMQ. To this end, we implement an open source network testbed with an automated execution of tests, which makes the entire environment easy to use, open to extensions and fully reproducible. The engineered testbed
interoperates with various MQTT protocol frameworks, whose utilization can be of interest to
both researchers and developers. We use two machine setups; a cluster of three virtual machines to measure the
performance when typical AMD64 nodes are used, as well as a cluster of three Raspberry Pis to
analyze the performance if these are replaced by less power-consuming ARM64 nodes.  To compare the performance of different broker implementations, we
leverage different testbed options for the configuration of its underlying network system, adjusting the parameters network delay, variance and packet loss. The broker performance is then benchmarked in terms of offset response time and message exchange latency for different payload sizes. The work is truly in progress, allowing us to analyze not only the performance, but also the choice of underlying hardware platform, such as whether the brokers are supported on different CPU architectures.

The rest of the paper is organized as follows. Section II presents the related work. Section III
describes the testbed and its configurations. Experimental benchmarking tests and results are presented
in Section IV. Section V concludes the paper and provides an outlook.

\section{Related work}

In the recent years, numerous efforts have been dedicated, by both the research community and the industry, 
to investigate performance, evolution and utilization of the MQTT protocol, as it still remains one of 
the most widely adopted M2M/IoT protocol solutions. This includes various studies that evaluate and compare different MQTT broker implementations based on their key features 
\cite{mishra2020use}. Many efforts have also been directed towards different benchmarking and evaluation methodologies for comparison of
MQTT broker performances in specific IoT scenarios \cite{gheorghe2020performance}. Common evaluations also include 
analyzing one or more MQTT broker implementations alongside the frameworks of other IoT suitable communication 
protocols \cite{herrnleben2021combench, 9090208}.
Paper \cite{mishra2018performance} compares the message delivery time for five implementations ActiveMQ, Bevywise MQTT, HiveMQ, Mosquitto and RabbitMQ, using a local deployment setup for the client side, while running the brokers in the cloud, reporting a very small difference in the obtained results.
An extensive evaluation of popular message brokers with MQTT support 
comparing Mosquitto, ActiveMQ, HiveMQ, Bevywise, VerneMQ, and EMQX  has been conducted in \cite{mishra2021stress}. The study focused on two testing environments; a local setup and a proprietary cloud solution, with Mosquitto outperforming others for the most of the observed metrics. Paper \cite{8783166} assesses the performance of Mosquitto and RabbitMQ in a Smart City scenario, showing the significant effect the choice of hardware platforms has on the performance. While there is no lack of studies of different aspects of MQTT communication in the cloud environments, particularly in terms of various  whitepapers provided by broker vendors, there are far less results in the edge context. In \cite{koziolek2020comparison}, authors compare distributed MQTT brokers (EMQX, HiveMQ and VerneMQ) in terms of scalability, with all three showing satisfying results. 

Similar to related works, we also focus on comparing commonly used open source MQTT broker implementations, but we focus on the choice of the solutions with official containerization support (whether on only AMD64 or both AMD64 and ARM64 nodes). In our case, we provide measurements within a network testbed (and not on cloud only, or local machines), which is a novel contribution.

\section{Testbed}

\begin{figure*}[htbp]
\centering
\hspace*{-0.7cm}\includegraphics[width=0.9\textwidth]{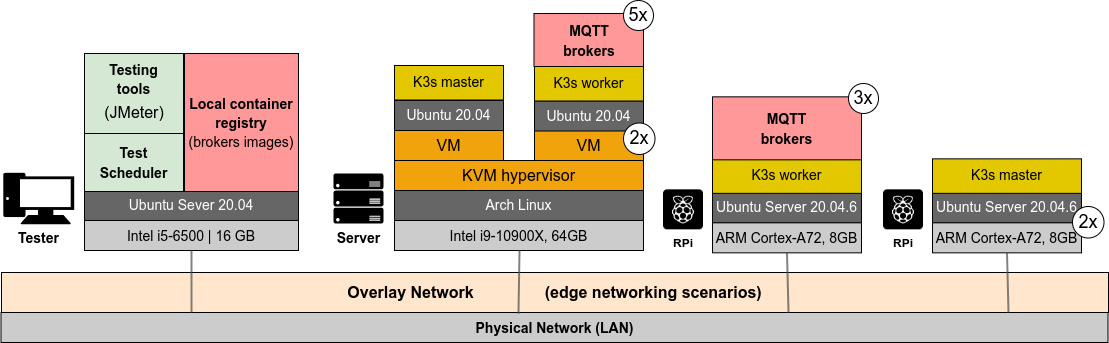}
\vspace{-0.3cm}
\caption{\label{fig:testbed}Testbed for experimental benchmarking of MQTT broker implementations}
\end{figure*}

Fig. \ref{fig:testbed} shows the reference network testbed that we engineered for experimental
benchmarking of MQTT broker implementations. The hardware configuration of the testbed mainly
consists of three Raspberry Pis (ARM Cortex A72, 8GB) and one server (Intel i9-10900X, 64GB), where
we deploy three virtual machines (4vCPU, 8GB). We differentiate between two testbed setups, one in
which MQTT brokers are deployed across moderately powerful AMD64 devices (VM setup) and the other
one in which they are deployed across more resource constrained ARM64 devices (RPi setup).
Furthermore, we employ a testing machine  (Intel i5-6500, 16GB), and a Gigabit Ethernet switch
through which all devices are interconnected.  The operating systems include Ubuntu version 22.04.2
for the tester, Ubuntu 20.04.6 for all Raspberry Pis (ARM64) and VMs (AMD64), and Arch Linux for the
server machine. It should be noted that we made use of a modified version of an open source
network testbed from  \cite{carpio2022benchfaas}. While the original testbed design was
engineered to benchmark serverless functions, we removed the serverless software framework and
reused its existing network and hardware abstractions. In this way, the present implementation
is independent and can be deployed and further developed as stand-alone open source MQTT broker testbed implementation.

Since the common way of deploying MQTT brokers is through containers, we deploy k3s - a lightweight
Kubernetes distribution - as a container orchestrator on top of the operating system. Across the
respectively used nodes (3 RPis or 3 VMs), a Kubernetes cluster is then formed with a single master
node and two worker nodes which handles the deployment and lifecycle management of the brokers. As
can be seen in Fig. \ref{fig:testbed}, all five brokers have official container images with AMD64
support available which are deployed on our VM setup, while only three of them can be deployed on
RPi setup. All nodes considered in each setup are connected through a Nebula\footnote{https://github.com/slackhq/nebula} overlay network with established virtual interfaces.
Finally, we have the tester device running a local container registry, testing tools and the test
scheduler. The local container registry contains the stable container images of the MQTT brokers for
the later deployment, while the main component of the tester - the test scheduler - is in charge of
executing the different benchmarking tests. For creating the tests, we use an MQTT plugin for the
JMeter load testing tool\footnote{https://github.com/emqx/mqtt-jmeter}. The test scheduler is also
in charge of orchestrating the testbed deployment, including the creation, provisioning and
destruction of required VMs, the setup of the overlay network and the configuration of network
parameters. Detailed instructions on the testbed’s usage and deployment as well as running the
various MQTT broker implementations over the VM or RPi setup are available on Github\footnote{https://github.com/michalke-it/benchmarking\_os\_mqtt\_brokers\_2023}. All results, software tools and code are fully reproducible and free and open source.

\subsection{Broker implementations}

Among the many MQTT broker options available, we opted for five open-source MQTT broker implementations written in different programming languages as depicted in Table \ref{tab:brokers} along with the officially supported protocol versions.

\begin{table}[htbp]
    \centering
    \caption{Proposed MQTT broker implementations}
    \label{tab:brokers}
    \begin{tabular}{lcccc}
        \toprule
        \textbf{Name} & \textbf{Language} & \textbf{ARM64} & \textbf{MQTT 3.1.1} & \textbf{MQTT 5.0} \\
        \midrule
        Mosquitto \footnote{https://mosquitto.org/} & C & supported & supported & supported \\
        EMQX\footnote{https://www.emqx.io/} & Erlang & supported & supported & supported \\
        RabbitMQ\footnote{https://www.rabbitmq.com/} & Starlark & supported & supported & - \\
        VerneMQ\footnote{https://vernemq.com/} & Erlang & - & supported & supported \\
        HiveMQ\footnote{https://hivemq.com/} & Java & - & supported & supported \\
        \bottomrule
    \end{tabular}
\end{table}

One of the selection criteria considered was the importance of ARM64 architecture support, which we
consider a crucial feature for future edge computing scenarios due to its high power efficiency.
For this reason we opted for brokers with official container images that support ARM64, which at the
moment includes only three implementations. Namely, these are the most popular and widely used
Mosquitto, the highly scalable EMQX, and RabbitMQ which was initially designed for the AMQP protocol
but also supports MQTT. In addition, we benchmark two other implementations that have been used in
distributed edge computing; HiveMQ and VerneMQ \cite{koziolek2020comparison}. Mosquitto, EMQX  and
RabbitMQ are analyzed on the RPi setup, while all five brokers are analyzed on the VM setup..

\subsection{Network configuration scenarios}

When it comes to the edge computing context, there is an additional dimension to consider besides potentially different CPU architectures and capabilities of hardware platforms; the QoS impacted by the different network parameters between all nodes. To explore potential performance differences between the brokers when this factor is considered, we run our tests over different network configuration scenarios, designed to approximate real-world scenarios. To do so, we
modify the latency, jitter and packet loss of the interconnections by applying the values to the
virtual overlay network interfaces.
The exact values for the whole system are listed in Table
\ref{tab:qos}, with each listed value representing the parameters that can be measured across the connection of any two nodes of the network, as well as the connection of the tester to any node. The values are defined based on the latency values measured in \cite{charyyev2020latency} for edge data centers and the scenarios defined in \cite{carpio2022benchfaas}.

\begin{table}[htbp]
    \centering
    \caption{QoS Parameters for Network configurations}
    \label{tab:qos}
    \begin{tabular}{lccc}
        \toprule
        \textbf{Scenario} & \textbf{Latency (ms)} & \textbf{Variance (ms)} &\textbf{Packet Loss (\%)} \\
        \midrule
        local & 0 & 0 & 0 \\
        optimal & 2.5 & 0.5 & 0.04 \\
        worst & 6.25 & 1.25 & 0.1 \\
        \bottomrule
    \end{tabular}
\end{table}
According to these QoS parameters, we define three network scenarios in which the broker
implementations will be benchmarked: local, optimal, and worst case. While the local scenario serves
as a baseline without any additionally introduced latency, variance or packet loss, the optimal
scenario represents what we consider an achievable best-case scenario for an edge computing
deployment with very low latency. The worst-case scenario on the other hand represents the highest
latency that could occur before a cloud deployment would provide better service quality. It has to be
noted that the listed parameters are applied per direction.

\subsection{Testing}
The testbed needs to be provided with configuration parameters specific to the infrastructure and tests. This is achieved through the customization of the
\textit{config.yml} file
\footnote{https://github.com/michalke-it/benchmarking\_os\_mqtt\_brokers\_2023}
for the parameter configuration, as well as through population of the \textit{performance-tests} folder with JMeter's test scripts. The important \textit{config.yml} settings include:
\begin{itemize}
    \item \emph{repository}: our local container registry that hosts the broker images
    \item \emph{ports}: the ports to be targeted for each test  
    \item \emph{tests}: the test(s) that should be ran in both setups (RP and VM)
    \item \emph{devices}: the IP addresses and login names of the devices that form RP and VM setup
    \item \emph{qos parameters}: specifies different latency, packet loss and jitter values for
    different network scenarios that are to be added to PM and VM network interfaces
    \item \emph{software}: the software versioning (Nebula, k3s, etc.)
    \item \emph{benchmark\_ip}: the IP address and ports to run the benchmarking tests against
\end{itemize}

After provisioning all devices, the testbed waits for all deployments to complete and then  proceeds to run JMeter's performance benchmarking tests. This procedure is then repeated for all the different sets of QoS parameters
specified in the configuration while modifying the MQTT communication parameters (number of publisher and subscriber threads, topics, messages to be published, timeouts etc.) accordingly every time before the testing scripts are executed, to
measure performance offset, response time and latencies for different message payload sizes. After configuration, the testbed waits or all deployments to complete and then we proceeds to run JMeter's benchmarking tests for performance, modifying MQTT communication parameters for testing scripts, which are to measure performance offset  response time and latencies for different message payload sizes. 

\section{Experimental Benchmarking}

\begin{figure*}[!t]
    \centering
    \subfloat[ Raspberry Pi
    setup]{\includegraphics[width=0.52\textwidth]{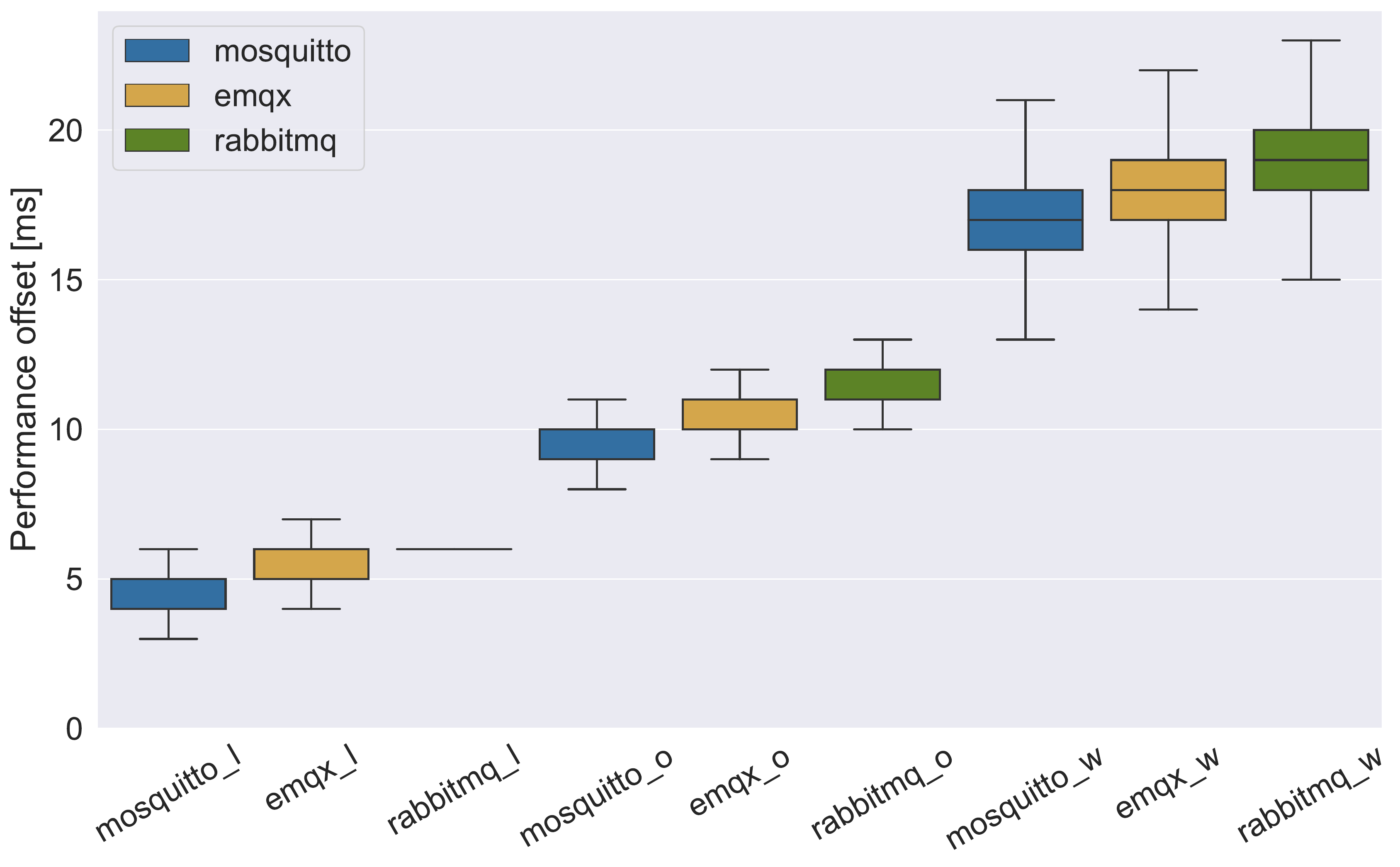}\label{rpi-overhead}}
    \hfil \hfil
    \subfloat[Virtual Machine setup]
    {\includegraphics[width=0.48\textwidth]{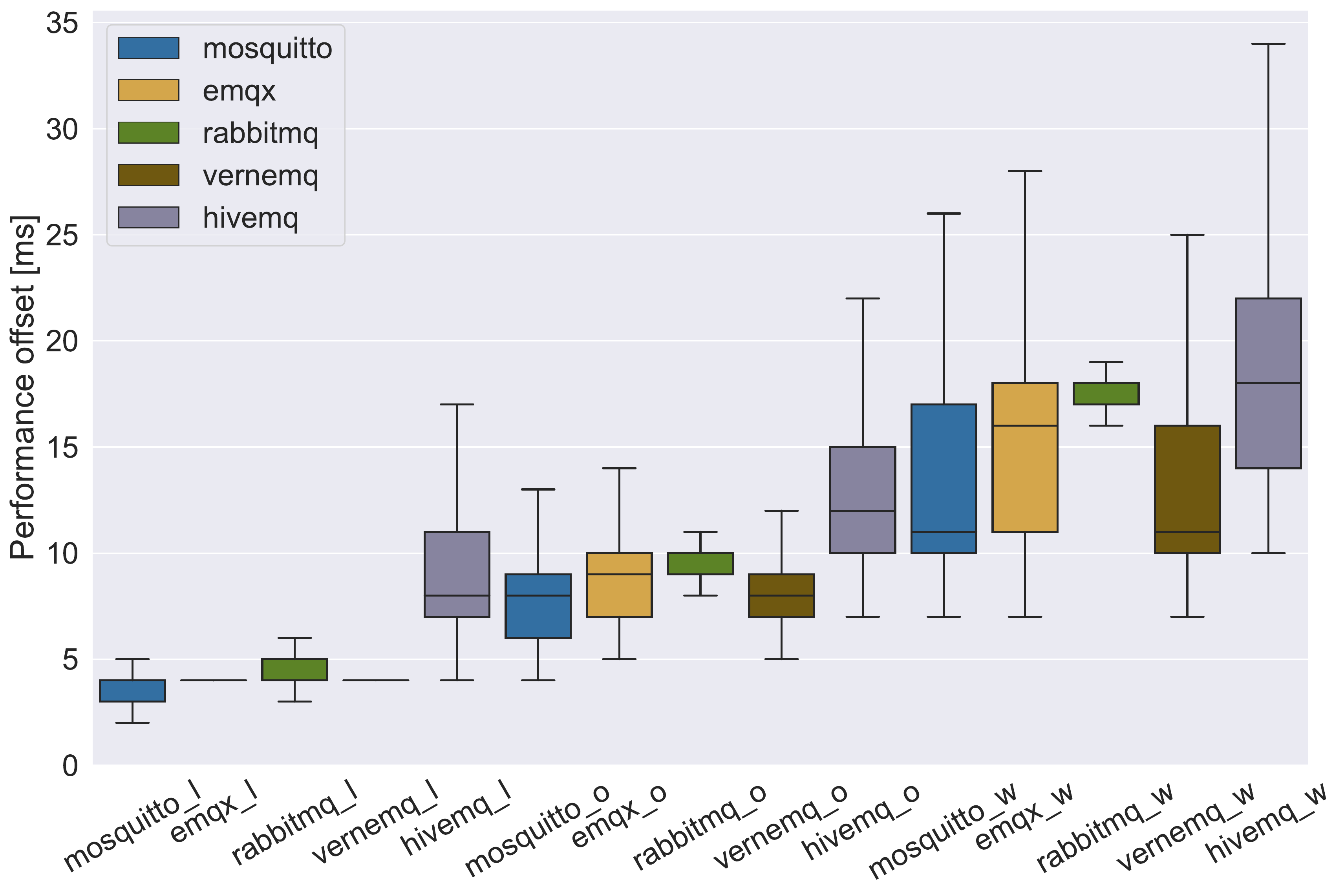}\label{vm-overhead}}
    \caption{Testbed performance offset}
\end{figure*}

In this section, we describe the benchmarking tests conducted as well as the measurements obtained.
We define two types of tests, one that measures the testbed performance offset and the other that
measures the payload size response time in terms of latency. The latter is calculated for the MQTT
publish/subscribe message exchange service, using MQTT protocol version 3.1.1. We calculate the
latency as the time elapsed from sending a message from the MQTT publisher to the MQTT subscriber.
As testing tool we use a JMeter MQTT plugin with customized test plans. We consider two hardware
setups for each test, i.e., VM and RPi, for three different network scenarios, i.e., local, optimal
and worst-case. In each test, we use one subscriber thread and adjust the number of publisher
threads, with each publisher thread sending a single message to the target broker. The published
messages are set to include timestamps, which are then subtracted from the corresponding timestamps
on the subscriber side. Each test is repeated at least 10 times to minimize the potential margin of
error and ensure reproducibility of the results. 

\subsection{Performance Offset}

This initial measurement serves to evaluate the performance
offset introduced when publishing a simple hello world string
to the MQTT brokers deployed on the testbed. Here, we
conduct two sets of measurements of the response time on
the subscriber side. In the first set, we analyze  broker implementations with official ARM64 container support (Mosquitto, EMQX and RabbitMQ). In the second set, we compare all five
of the initially selected implementations (Mosquitto, EMQX, RabbitMQ, HiveMQ
and VerneMQ). On the publisher side,
we configure a utilization value of 100 threads, each sending the
same message with an interval of 250ms. The QoS of both,  i.e.,
the subscriber and the publisher, is set to the reliable level
1, guaranteeing a successful message transfer to the broker.
It should be noted that neither the authentication nor the security mechanism was implemented.

For the first set of measurements, the response time values for the brokers running on the RPi setup
are shown in Fig.\ref{rpi-overhead} with boxplot graphics showing the median value for each of the
broker as well as variations in the obtained data. Here, we can see that the median value for the
performance offset tests is around 5 ms in the local scenario, 10 - 12 ms in the optimal scenario,
and 18 - 19 ms in the worst-case scenario, for all implementations. The figure also shows that the
boxplots are almost of the same size for each network scenario with low variation of the values.
While Mosquitto seems to preform slightly better, the comparable delays in all cases are very low
and can be considered negligible.

For the second set of measurements, the response time values for all five brokers running on the VM
setup are shown in Fig.\ref{vm-overhead}. The median value for performance offset is around
4.8 - 8 ms (local), 8 - 13 ms (optimal), and 11 - 18 ms (worst-case scenario), for all implementations. This case however shows higher variation of the values, particularly in the worst-case scenario, and
especially for HiveMQ broker, which shows the highest deviation of all brokers, even under
comparably good conditions. This behavior becomes more pronounced the worse the networking
conditions become, notably exceeding the other brokers' deviations while maintaining a similar
median response time. The opposite effect can be observed with RabbitMQ broker, which shows
slightly elevated median response times while maintaining significantly more predictable response
times than the other brokers. When comparing the median values of the three broker implementations
that have been tested on both setups in in Fig.\ref{rpi-overhead} and Fig.\ref{vm-overhead}, we can
also see a slightly lower response time by roughly one ms which can be considered negligible. This
difference can also be traced back to the slight performance advantage that the VMs have over the
RPis since the VM's inter-node traffic does not have to pass through a physical network medium but
remains in the same physical device instead.

\subsection{Payload size}

\begin{table*}[!t]
    \begin{center}
     \setlength\tabcolsep{1.2pt}
     \caption{Payload size: latency median and IQR - Mosquitto, EMQX and RabbitMQ}
     \label{table:rpi_vm_payload}
    \begin{tabular}{llccccccccc }
        \toprule
   
     &  & \multicolumn{3}{c}{\textbf{Mosquitto}} & \multicolumn{3}{c}{\textbf{EMQX}} & \multicolumn{3}{c}{\textbf{RabbitMQ}}\\[0.5ex] 
     &  & local    & optimal   & worst   & local  & optimal & worst & local   & optimal   & worst   \\ 
     \midrule 
     \multirow{2}{*}{1KB}   & RP  & 5.39 - 1.00 & 10.67 - 1.00 & 18.60 - 2.00& 6.04 - 0.25 &
     11.54 - 1.00 & 18.93 - 2.00 & 7.00 - 1.00 & 12.00 - 1.00  & 19.87 - 2.25 \\ [0.5ex] 
       & VM & 3.94 - 0.00 & 8.59 - 2.00  & 15.23 - 6.00 & 4.24 - 0.00 &  9.14 - 3.00 & 15.25 -
       7.00  & 4.83 - 1.00 & 9.65 - 3.00 & 15.59 - 6.00   \\ [0.5ex] \hline
                                   
\multirow{2}{*}{10KB} & RP & 10.68 - 1.00  & 15.91 - 1.00  & 24.90 - 2.00  & 11.44 - 2.00 & 16.56
- 2.00 & 24.71 - 2.00 & 13.16 - 2.00 & 17.73 - 1.00  & 26.15 - 2.00 \\ [0.5ex] 
   & VM & 4.33 - 0.00 & 8.83 - 2.00  & 16.85 - 7.00  & 4.66 - 1.00 & 9.43 - 2.00  & 16.67 - 7.00
   &  5.76 - 1.00 & 9.85 - 1.00 & 17.25 - 6.00   \\ [0.5ex]  \hline  

\multirow{2}{*}{1MB}   & RP & 272.48 - 32.00  & 332.42 - 48.00  & 643.17 - 181.25 & 274.11 - 35.00&
333.04 -49.00 & 641.02 - 193.75  & 297.74 - 53.00 & 379.31 - 55.00 & 679.16 - 167.00 \\ [0.5ex] 
    & VM & 66.27 - 13.00 & 153.96 - 75.00 & 437.10 - 388.25 & 65.2 - 14.00  & 149.60 - 69.25  &
    418.81 - 356.50 & 71.97 - 13.00 & 161.38 - 64.25 & 429.83 - 379.50  \\ [0.5ex] \hline
    \end{tabular}
\end{center}

\label{table:results_123}
    \end{table*}

    For the payload tests, we also configure a utilization of 100 threads on the publisher side, sending messages of different payload sizes with an interval of 250ms. We modify the message size by defining strings of different fixed lengths, considering 1KB, 10KB and 1MB sizes. Next, we measure latency on the subscriber side for each payload size and for each broker implementation in all three network scenarios, for the VM and for the RPi setup. The QoS of both, the subscriber and publisher, remains on level 1. 

    For the first set of broker implementations tested on both setups, i.e., RPi and VM, the median latency
    values and interquartile range (IQR), which shows how spread the boxplot is, are shown in Table
    \ref{table:rpi_vm_payload}, with their values in milliseconds. When sending a 1KB message, the
    resulting median values in both setups are very close in all three network scenarios, with
    negligible differences between the broker implementations. Even the differences in terms of latency
    values between the local and worst-case scenario (e.g., Mosquitto, local median 5.39 ms and worst
    case median 18.6 ms) are quite low. If we compare these median values with the ones obtained as
    performance offset and shown in Fig.\ref{rpi-overhead} (e.g., approximately 4.8 ms and 17 ms median
    values for local and worst case scenario for Mosquitto) we can see that the 1KB payload size
    introduces a very low additional latency. The IQR values also are in a low range, with slightly
    higher variance for the worst case network scenario of RabbitMQ and EMQX
    implementations. Similar trends continue for 10KB message size, with the expected increase of
    latency in all cases compared to 1KB message size. This increase remains in the range of
    approximately 5 ms in all scenarios and all implementations.

\begin{figure}[!t] \label{vmrp}
    \centering
 {\includegraphics[width=1.0\columnwidth]{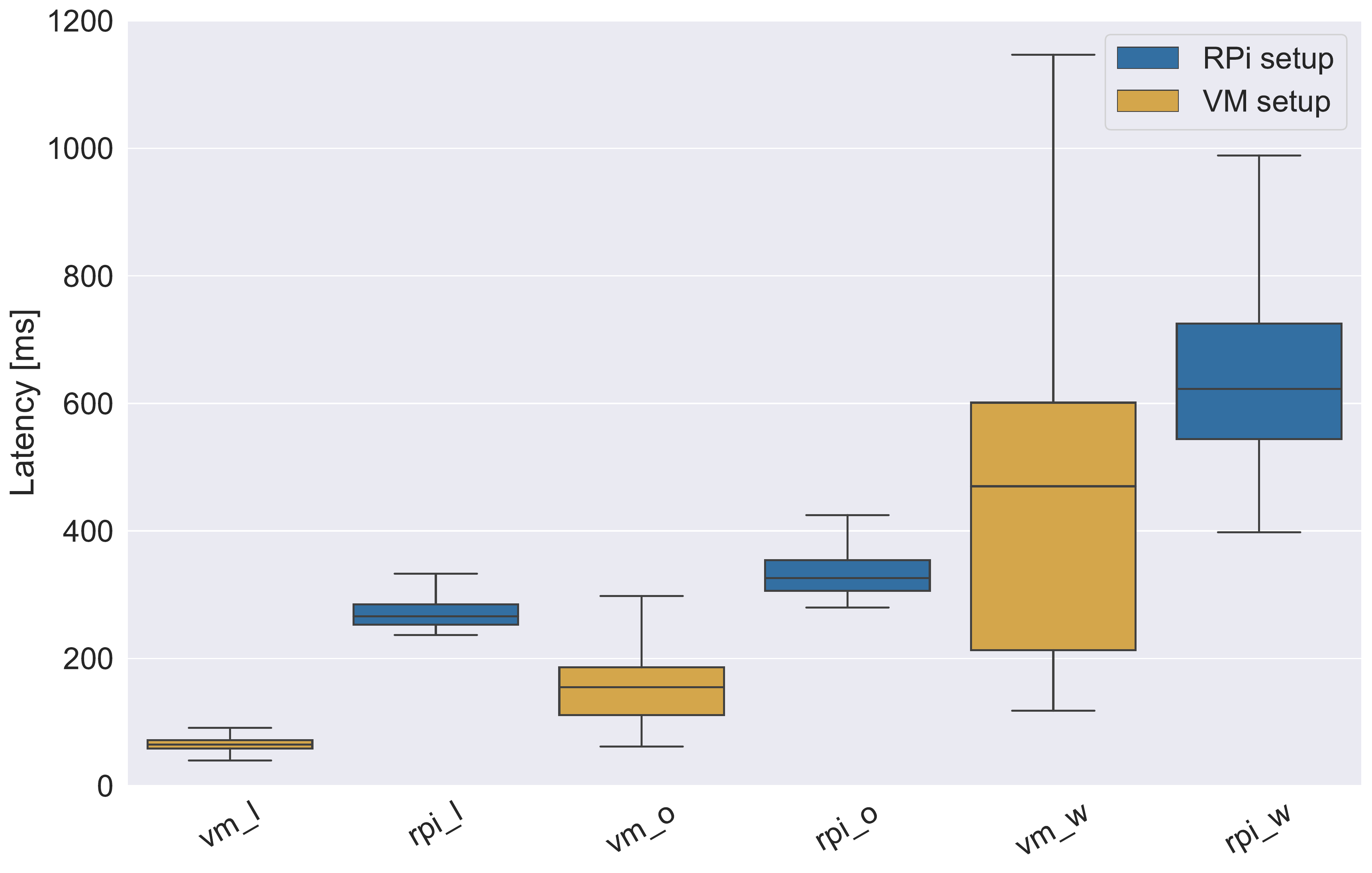}} 
\caption{Payload size 1MB: RP setup vs VM setup}
\end{figure}

Moving to 1MB message size, we can observe significant performance differences. First, when
comparing the same broker implementations on the VM and RPi setup, we can observe that in each
broker implementation and in each network scenario, VMs preform better with noticeably lower
median response times, as well as lower variance. The difference in latency between these two
hardware setups is approximately 200ms for all implementations, which can be a limiting factor in
low-latency demanding IoT applications, where MQTT is often the protocol of choice. 
On the other
hand, when comparing the median response times for the same scenario on the RPi setup, e.g. 643.17
ms (Mosquitto), 641.02 ms (EMQX) and 679.16 ms (RabbitMQ) for the worst-case scenario, we can see
that the broker implementation itself does not play a key role in the overall results. The same can
be said for VM setup. Unlike for the small payload sizes, in the case of 1MB, the IQR range is
much higher, indicating a higher variance of the obtained values. 

The effect that the setup has on which the brokers were deployed is also visually shown on the example of RabbitMQ broker in  
Fig. 3 for  1MB message payload size. As it can be seen, the difference in the response time on VM and RP setup is noticeable in each of the three network scenarios, and it plays almost the same role as the degradation of QoS parameters in the scenarios themselves.   

\begin{table}[htbp]
    \centering
    \caption{Payload size: latency median and IQR - HiveMQ and VerneMQ}
    \label{table:vm_payload}
    \begin{tabular}{llccc}
        \toprule 
     &   & \multicolumn{3}{c}{\textbf{HiveMQ}} \\[0.5ex] 
     &   & local  & optimal  & worst  \\ 
     \midrule
    \multirow{1}{*}{1KB}    & VM & 9.70 - 5.00 & 13.46 - 5.00 &  18.40 - 7.00 \\[0.5ex]  \hline     
    \multirow{1}{*}{10KB} & VM  & 9.95 - 4.00   & 13.94 - 5.00 &  19.75 - 8.00  \\ [0.5ex]  \hline 
    \multirow{1}{*}{1MB}   & VM & 125.00 - 24.00 & 194.72 - 73.00 & 374.80 - 320.00  \\ 
     \bottomrule
     \\[0.5ex] 
     & & \multicolumn{3}{c}{\textbf{VerneMQ}} \\[0.5ex] 
     & & local   & optimal & worst  \\ 
     \midrule
     \multirow{1}{*}{1KB}    & VM & 4.37 - 1.00  & 8.30 - 3.00  & 13.64 - 6.00  \\ [0.5ex]  \hline                          
     \multirow{1}{*}{10KB} & VM   & 5.18 - 1.00  & 9.23 - 2.00 & 15.23 - 7.00 \\ [0.5ex] \hline
     \multirow{1}{*}{1MB}   & VM  & 76.25 - 12.00 & 144.85 - 71.00  & 345.64 - 292.50  \\
      \bottomrule
    \end{tabular}

    \end{table}

    The median latency values and IQR for the remaining broker implementations, which are tested only on
    the VM setup, are shown in Table \ref{table:vm_payload}. For 1KB message we can see the same trend
    as in the previously discussed implementations, with the median latency values slightly higher in
    the local network scenario for HiveMQ implementation. If we compare these median values with the
    ones obtained as performance offset and shown in Fig.\ref{vm-overhead} (e.g., approximately 8 ms and
    18 ms median values for local and worst-case scenario for HiveMQ) we can again see that 1KB
    payload size introduces negligible latency. The same trend continues for 10KB message size, with
    the increase of latency compared to 1KB message being less than 1 ms. Moreover, this result can be observed in all five broker implementations tested on VM setup. As previously, the more significant deteriorations in performances of both median latency value and IQR range can be noticed with 1MB payload size. However, if we now compare all median response times for the same scenario and the VM setup, e.g., 153.9 ms (Mosquitto), 149.6 ms (EMQX), 161.38 ms (RabbitMQ), 194.72ms (HiveMQ) and 144.85 ms (VerneMQ) for optimal case scenario, we can see that the differences are not very pronounces, with HiveMQ performing slightly worse than other implementations. 
    However, if we consider the worst-case scenario on the VM setup, we can see a significant difference
    between the two additional brokers VerneMQ and HiveMQ and the remaining brokers. Namely, their mean
    values of 345.64 ms (VerneMQ) and 374.80 ms (HiveMQ) are considerably lower than 429.83 ms
    (RabbitMQ), 418.81 ms (EMQX), and 437.10 ms (Mosquitto) scored by the remaining brokers. This is
    especially interesting given that HiveMQ preformed worse than the others when using
    smaller payloads.

\section{Conclusion and outlook} \label{concl} 
We experimentally benchmarked the performances of five open-source MQTT Broker implementations. To this end, we engineered a network testbed in an edge computing context with constrained devices. When analyzing performance offset of the testbed itself, we found comparable response times for all brokers, when deployed in the same network scenarios and on the same hardware platform setup. In cases of small message payload sizes (up to 10KB), with degradation of network conditions,  a small latency increase has been noticed for each of the broker implementations. In that case under the same conditions, the latency differences between them could be considered negligible. Once the message payload size increased to 1MB, the performance of all implementations in all network scenario deteriorates  in both the median latency and its deviation. However, the largest performance differentiator has been in the testbed's hardware setup itself, with Virtual Machine (VM) setup resulting in much lower response times than the Raspberry Pi setup. 

The experiments were conducted as a support in the decision making process for the most suitable broker implementation. 
We conclude that this decision has to be made with strong consideration for the network environment and use cases, e.g., how big the typical message sizes are. In future work, the performance should also be explored in relation to the scalability throughput and the number of concurrent
publishers and subscribers. In addition, the engineered open source network testbed can be further extended to include and benchmark
other MQTT broker implementations, as well as to integrate other communication protocol broker frameworks, which can then be evaluated with  additional tests and scenarios.

\section*{Acknowledgment}

This work was partially supported by EU HORIZON research and
innovation program, project ICOS, Grant Nr. 101070177,  and by  European Commission under the H2020-952644 contract for project FISHY.

\bibliographystyle{IEEEtran}

\bibliography{gllobecom2023bibliography}

\end{document}